# Determining The First Order Character of La(Fe, Mn, Si)$_{13}$

Milan Bratko, Edmund Lovell, A. David Caplin

*Blackett Laboratory, Imperial College London, Prince Consort Road, London, SW7 2AZ, United Kingdom*

Vittorio Basso

*Istituto Nazionale di Ricerca Metrologica, Strada delle Cacce 91, 10135, Torino, Italy*

Alexander Barcza, Matthias Katter

*Vacuumschmelze GmbH & Co. KG, Hanau, Germany*

Lesley F. Cohen

*Blackett Laboratory, Imperial College London, Prince Consort Road, London, SW7 2AZ, United Kingdom*

Definitive determination of first order character of the magnetocaloric magnetic transition remains elusive. Here we use a microcalorimetry technique in two modes of operation to determine the contributions to entropy change from latent heat and heat capacity separately in an engineered set of La(Fe, Mn, Si)$_{13}$ samples. We compare the properties extracted by this method with those determined using magnetometry and propose a model independent parameter that would allow the degree of first order character to be defined across different families of materials. The microcalorimetry method is sufficiently sensitive to allow observation of an additional peak feature in the low field heat capacity associated with the presence of Mn in these samples. The feature is of magnetic origin but is insensitive to magnetic field, explicable in terms of inhomogeneous occupancy of Mn within the lattice resulting in antiferromagnetic ordered Mn clusters.

## I. INTRODUCTION

La(Fe, Si)$_{13}$ based compounds are promising candidates for solid state magnetic cooling, exhibiting a large magnetocaloric effect (MCE) associated with a first order metamagnetic phase transition above the Curie temperature, $T_C$, and are attractive due to being made up mainly of highly abundant materials as well as potentially offering modest magnetic and thermal hysteresis. The $T_C$ is tunable by substitution onto the Fe site. $T_C$ increases with increasing Si content, for example,[1,2] and the sharp features observed in magnetization for low Si concentrations, broaden as the material moves from a first order to a continuous phase transition. Strongly first order materials show thermal and magnetic hysteresis, which limits the available entropy and adiabatic temperature changes available in the refrigeration cycle, and also introduces loss.[3,4,5] $T_C$ can also be shifted to near room temperature by hydrogen absorption while sustaining the large MCE.[6,7] Partial replacement of Fe by other transition metal elements such as Mn, Co, Cr and Ni, and interstitial atoms such as B, C, N and H have been explored both experimentally and theoretically.[8] Most commonly, a combination of Mn substitution, Si composition and hydrogenation is used to optimise the magnetocaloric properties, bringing the transition as close to first order as possible whilst engineering a range of $T_C$ so that a cascaded set of solid state refrigerants can be employed, for refrigeration applications with a useful range of working temperatures.[9,10,11,12] Previously, the LaFe$_x$Mn$_y$Si$_{13-x-y}$ system was studied as a function of Mn content. It was found that $T_C$ decreased monotonically with increasing Mn concentration from 188 to 127 K, and the saturation magnetization, $m_{sat}$, decreases from 23.9 $\mu_B$/f.u. to 22.2 $\mu_B$/f.u. respectively, as $y$ increases from 0 to 0.35.[9] The decline of $m_{sat}$ was found to be faster than simple magnetic dilution. This may have two causes. One is that the magnetic moment per Fe atom is reduced due to the change of Fe chemical environment caused by the Mn substitution. The other is that the Mn atoms carry magnetic moment that couple antiparallel to the Fe moments. The latter has been recently confirmed theoretically.[8]

La(Fe, Si)$_{13}$ is an itinerant ferromagnet, showing a critical point, $T_{crit}$, in its H-T phase diagram. At temperatures and fields below $T_{crit}$, the transition between paramagnet and ferromagnet is first order in character, showing thermal and magnetic hysteresis. Above $T_{crit}$, the transition shows the signatures of a continuous phase transition, no hysteresis and a significantly broadened transition. There are a number of models based on the Landau expansion of the free energy, used to parameterize the order of the transition: the Banerjee criteria,[13] the Arrot plot,[14] the Bean Rodbell model[15] and its extensions,[16] and for itinerant systems, spin fluctuation theory.[17,18] However, most of these models require a number of parameters to be defined including those related to real materials, such as inhomogeneous spread of $T_C$, and clustering.[19] It is difficult to compare first order character between materials when different models apply to different types of magnetic systems (local and itinerant magnetism). Although hysteresis is considered to be a signature of first order character, we have previously shown that there are extrinsic contributions to hysteresis,[20,21] and that the relationship between latent heat and hysteresis is different for different material families.[22] Consequently a direct measure of the degree of first order character is lacking.

Recently, the tuning of $T_{crit}$ was explored in a series of the La(Fe, Mn, Si)$_{13}$-H$_{1.65}$ from the characteristic changes in heat capacity.[23] In this paper we consider the matching family of La(Fe, Mn, Si)$_{13}$ materials (that is without the hydrogenation). We consider the order of the transition by extracting the latent heat explicitly and show how it is suppressed in applied magnetic field as the critical point is approached. We show the influence on this behavior on one sample that has been hydrogenated. For a representative set of samples we compare the latent heat in field with the information that can be extracted from magnetization using the Clausius-Clapeyron equation and Maxwell relations,[24] and use this to define a model independent parameter of first order character, $\Omega$. The ac calorimetry measurements reveal an additional feature which we interpret as being due to antiferromagnetic regions in the sample of the order of 20% of the total volume due to Mn clusters.

## II. EXPERIMENTAL METHOD

### A. Samples

La(Fe, Mn, Si)$_{13}$ alloys with variable Mn content were prepared by powder metallurgy technique and hydrogenated as described in Barcza *et al*.[10] Master alloys were prepared by vacuum induction melting followed by mechanical milling steps to produce fine powders. The composition of each alloy was adjusted by blending master alloys with elemental powders. Compaction of the powder blends was performed by cold isostatic pressing. The green bodies were vacuum sintered at around 1100 °C followed by an annealing treatment at 1050 °C.[25] Hydrogenation was performed on a granulate material with a particle size less than 1 mm by heating to 773 K in argon. At 773 K argon was replaced with hydrogen followed by a slow cool to room temperature. The compositions are summarised in table I.

TABLE I. Summary of the $T_C$ and compositions of the series of LaFe$_x$Mn$_y$Si$_z$ compounds studied here.

| Sample | A | B | C | D | E | F | G |
|---|---|---|---|---|---|---|---|
| $T_C$ [K] (w/ H) | 269 | 283 | 293 | 313 | 323 | 333 | 343 |
| $T_C$ [K] (w/o H) | 110 | 131 | 142 | 150 | 158 | 168 | 173 |
| x (Fe) | 11.22 | 11.33 | 11.41 | 11.49 | 11.58 | 11.66 | 11.74 |
| *y* (Mn) | 0.46 | 0.37 | 0.30 | 0.23 | 0.18 | 0.12 | 0.06 |
| *z* (Si) | 1.32 | 1.30 | 1.29 | 1.27 | 1.25 | 1.23 | 1.20 |

## B. Magnetometry

All magnetization measurements were performed using a Quantum Design PPMS VSM option with external magnetic field up to 9 T. The isothermal entropy change, *ΔS*, was estimated from isothermal magnetization measurements using the Maxwell relation:

$$\left(\frac{\partial S}{\partial B}\right)_T = \left(\frac{\partial M}{\partial T}\right)_B, \qquad (1)$$

where *M* is the magnetization and *B* is the magnetic flux density, which we assume to be equal to $\mu_0 H$. In the vicinity of a hysteretic first order phase transition (FOPT), a measurement protocol consistent with Caron *et al.*[26] has been adopted in order to avoid non-physical overestimates of the isothermal entropy change.[27]

In order to estimate the latent heat contribution to the total entropy change from the magnetization data we have used the Clausius-Clapeyron equation:

$$\Delta S_{CC} = -\Delta M \mu_0 \frac{dH_C}{dT}, \qquad (2)$$

where *ΔM* is the change in magnetization at the FOPT and $\mu_0 dH_C/dT$ is the slope of the phase line of the FOPT.

## C. Microcalorimetry

Microcalorimetry measurements were performed using a commercial Xensor SiN membrane chip (TCG-3880) adapted to operate either as an ac calorimeter[28] or as a quasi-adiabatic temperature probe,[29] in a cryostat with temperature range 5-293 K and an external magnetic field up to 8 T. The sample is a fragment of the bulk, typically ~100 µm with mass of the order of few µg. For an accurate determination of mass, the fragments were measured in the magnetometer and the saturation magnetization of the ferromagnetic state was compared with bulk samples of known mass.

In the ac measurement a modulated power is applied to the sample and the heat capacity is determined from the phase and amplitude of the resulting temperature oscillations, which are measured using a lock-in amplifier. Thus, the technique measures only reversible changes in heat capacity and the latent heat is ignored because of the hysteresis associated with it. The ac heat capacity measurement is absolute: however, the sensitivity of the thermopile used to measure the temperature oscillations has to be calibrated. For this purpose the temperature dependence of the heater resistance is used as a reference measure of temperature. Nevertheless, due to the finite thermal resistance between the heater and the sample, the heater is always hotter than the sample and a fixed correction factor has to be applied to the thermopile sensitivity. The correction factor can be determined by comparing field induced entropy changes estimated from the ac heat capacity data and magnetization measurements. This was performed well-away from the first order phase transition, where both techniques should work reliably and produce comparable estimates.

Figure 1(a) shows the heat capacity measured as a function of temperature. The heat capacity can be used to calculate entropy change, *ΔS*. For a field variation from 0 T to $\mu_0 H$:

$$\Delta S(T) = \Delta S(T_{ref}) + \int_{T_{ref}}^{T} \frac{C_{p,\mu_0 H}(T) - C_{p,0}(T)}{T} dT, \quad (3)$$

where the reference entropy change at $T_{ref}$ can be obtained from magnetization measurements. The zero field and in-field heat capacity values used are both from either cooling or heating curves.

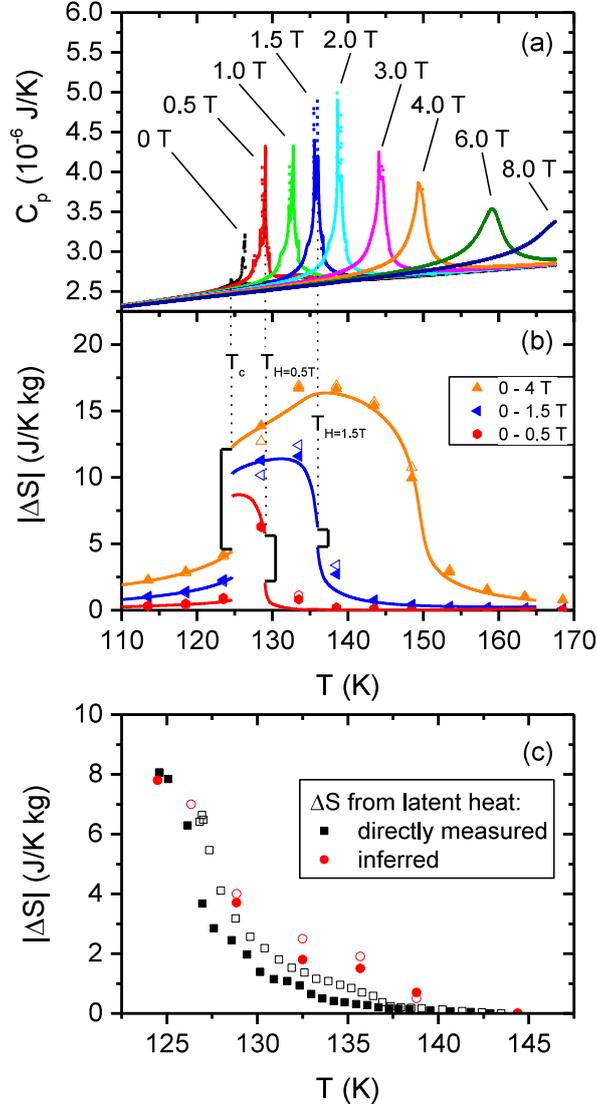

FIG. 1. (a) The ac heat capacity of sample B as measured in the microcalorimeter. (b) These entropy changes calculated from the ac heat capacity on cooling (lines) exclude the latent heat contribution and therefore require offsetting above $T_C$ and $T_H$ (the latter varies based on the upper field limit) in order to fit the total entropy change as estimated from the magnetization measurements (symbols). The manually fitted offsets (indicated by the brackets) offer an indirect measure of the entropy change associated with the latent heat. (c) Directly measured latent heat compared with that inferred from ac heat capacity data. Throughout, solid lines and symbols correspond to cooling or field application, dashed/open lines and symbols correspond to heating or field removal.

Since the ac calorimeter ignores latent heat and the associated entropy change, the result of eq. 3 becomes a convolution of the total entropy change and the latent heat released at $T_C$ (in 0 T) and/or $T_H$ (in $\mu_0 H$), once the heat capacity curves inside the integral cross their respective FOPTs. As explained in Morrison et al.,[30] this can be used to estimate the latent heat indirectly by comparison with entropy change estimates from magnetization measurements.

In order to reflect the total entropy change above $T_C$, eq. 3 can be modified to:

$$\Delta S(T_C < T < T_H) = \Delta S(T_{ref}) + \int_{T_{ref}}^{T} \frac{C_{p,\mu_0 H}(T) - C_{p,0}(T)}{T} dT + \Delta S_{LH}(T_C), \qquad (4)$$

and

$$\Delta S(T > T_H) = \Delta S(T_{ref}) + \int_{T_{ref}}^{T} \frac{C_{p,\mu_0 H}(T) - C_{p,0}(T)}{T} dT + \Delta S_{LH}(T_C) - \Delta S_{LH}(T_H), \qquad (5)$$

where $\Delta S_{LH}$ is the entropy change associated with the latent heat released at $T_C$ and $T_H$ for temperature sweeps in 0 T and $\mu_0 H$, respectively. $\Delta S_{LH}(T_C)$ and $\Delta S_{LH}(T_H)$ can be used as fitting parameters for the total entropy change estimated from the magnetization measurements and yield indirect estimates of the latent heat. An example of this fitting procedure is shown in figure 1(b).

The microcalorimeter enables also a direct measurement of the latent heat in the quasi-adiabatic temperature probe setup which relies on the instantaneous release of latent heat as the sample is driven monotonically through the first order phase transition either by applying magnetic field or changing the temperature. The release of latent heat results in a sharp change of temperature of the sample (and addenda). This is recorded as a sharp spike in thermopile voltage with an exponential decay as the latent heat diffuses to bath. The temperature spike can be described by:

$$\Delta T = \frac{Q_{LH}}{C} e^{-\frac{C}{G}t}, \qquad (6)$$

where $Q_{LH}$ is the latent heat released, $C$ is the heat capacity of the sample and the addenda, and $G$ is the thermal conductance between the sample and the bath. In order to maximise the reproducibility of the measured peak height for a given amount of latent heat, the time constant is lengthened by evacuating the sample space to below $4\times10^{-2}$ mBar and thus reducing the thermal link to bath.

In the original measurement setup,[30] the peak height was used as the measure of latent heat, calibrated by a reference heat pulse of known energy from a local heater. This approach assumes that $C$ does not vary significantly between the measurement and the calibration. Nevertheless, in figure 1(a) it can be seen that in the samples studied here the background heat capacity may vary by as much as 100% at the first order phase transition when the latent heat is released. For this reason we have considered the area of the peak as a more reliable measure of latent heat as opposed to the peak height – the integral of equation 6 yields $Q_{LH}G$, where $G$ can be expected not to vary with applied field and vary only slightly over a small temperature range. Furthermore, the integration approach simplifies the data analysis in samples where a cascade of overlapping peaks is observed, as the successive peaks superimpose linearly and the whole cascade can be simply integrated.

This approach has been validated by performing a calibration in zero field away from a phase transition and at the same temperature in-field, in the vicinity of the heat capacity peak. While the height of the calibration peak varied significantly, the area of the calibration peak remained unaffected by the change in background heat capacity.

Figure 1(c) shows the thus evaluated directly measured latent heat compared with the latent heat inferred indirectly from the ac heat capacity data. The two measurement are in good agreement, thus validating the measurement method as well as our approach to separate the first order contribution to the total entropy change.

## III. RESULTS AND DISCUSSION

In figure 2 we show the changes of heat capacity and latent heat in sample E. When the system approaches the critical point, the latent heat drops to zero and characteristically, we see an increasing peak in the ac heat capacity.[31,32] Thus, while the total entropy change maintains a plateau-like shape typical for a first order phase transition, the first order contribution gradually decreases. After the transition becomes fully second order the peak in heat capacity broadens and subsides.

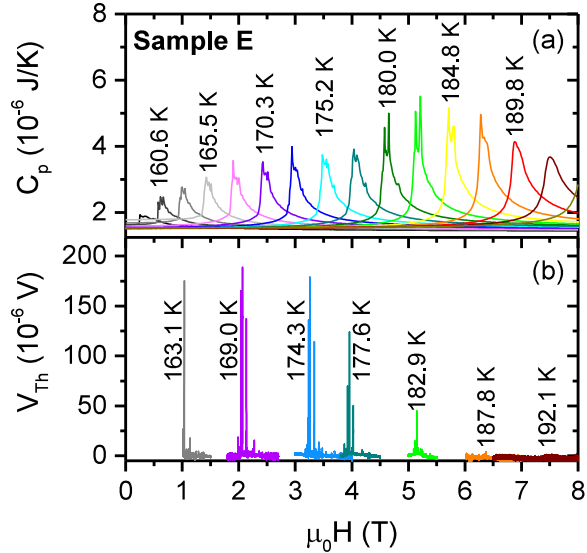

FIG. 2. AC heat capacity (a) increases with field/temperature while the latent heat signal (b) diminishes as the metamagnetic phase transition approaches criticality and becomes second order.

In order to evaluate the critical point a measure often used is the point of vanishing thermal/field hysteresis.[23] Figure 3 shows the $T_C$ and the $T_{crit}$ across the series using this method. The $T_{crit}$ has been identified as the point of vanishing hysteresis in the specific heat measurement for the hydrogenated samples (as reported in Basso *et al.*[23]) and in the magnetization measurements for the samples without hydrogen. It can be observed in figure 3(a) that $T_C$ changes systematically with introduction of Mn and that the same is true of the hydrogenated samples with their much higher $T_C$. The variation of $T_C$ with Mn is not greatly affected by hydrogenation, but both Mn and hydrogen significantly tune the critical point. Indeed in this sample series, the temperature separation between $T_C$ and the $T_{crit}$ could be used as a measure of first order strength of the transition. It can be seen that both Mn and H systematically weaken the first order character (i.e. $T_{crit}$ approaches $T_C$).

Figure 3(b) shows the phase lines for the dehydrogenated series and figure 3(c) shows the derivatives $\mu_0 dH_C/dT$. It has been discussed elsewhere,[33,34,35] that there is an optimum value of $\mu_0 dH_C/dT$ to achieve maximum entropy change, where it is assumed that $\mu_0 dH_C/dT$ is field invariant, which is clearly not the case here. We return to this point later. Using the direct latent heat measurement we can obtain significantly more detail on how the critical point is approached across the series and under the influence of hydrogen. We focus on samples B, E, G in the state without hydrogen and consider the effect of hydrogenation on sample B.

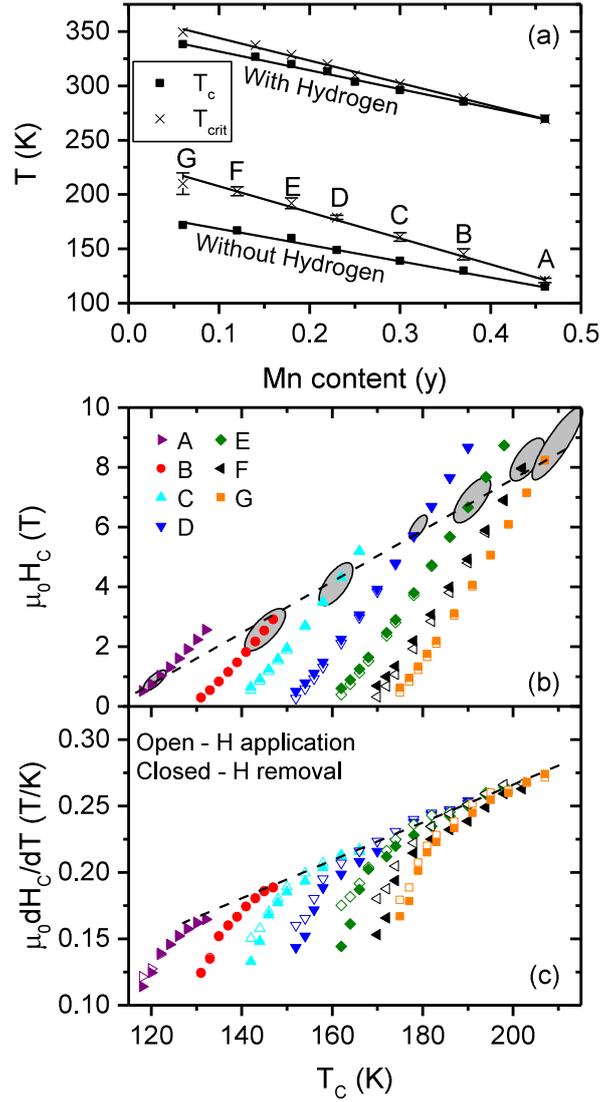

FIG. 3 (a) $T_C$ and $T_{crit}$ as a function of Mn doping in the samples with and without hydrogen. The data set with hydrogen was taken from Ref. 23. Phase lines (b) and their slope (c) derived from the bulk magnetization data. The critical point was determined from the point where the phase lines on field application and removal converge. The shaded areas in (b) indicate the uncertainty in the critical point temperature.

Figure 4(a-c) compares the directly measured latent heat contribution to the entropy change and the total entropy change estimated from bulk magnetization measurement for 0 to 1.5 T and 0 to 8 T (using the Maxwell relations, eq. 1). The latent heat contribution to the total entropy change decreases with increasing Mn content as the $T_C$ and $T_{crit}$ are brought closer.

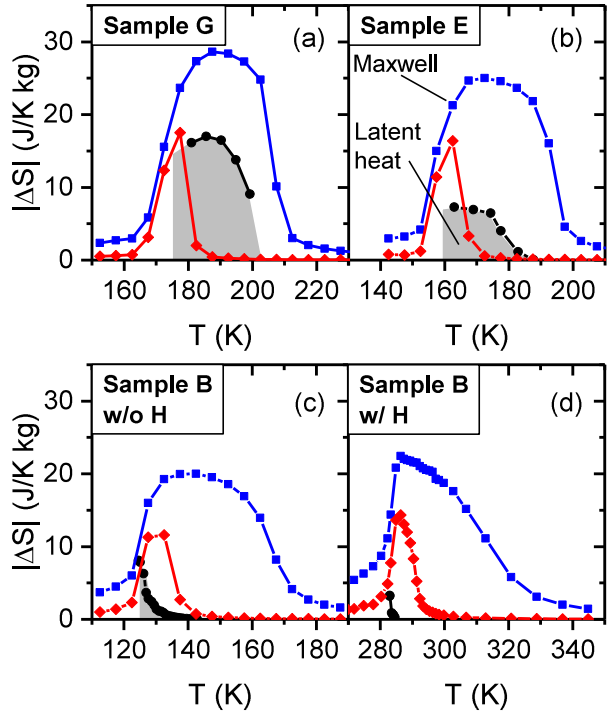

FIG. 4. Directly measured latent heat contribution to entropy change (black circles) at the FOPT compared with the total entropy change estimated from magnetization measurements for a magnetic field variation of 0 to 1.5 T (red diamonds) and 0 to 8 T (blue squares). All data correspond to magnetic field application/cooling.

In the most strongly first order sample, *G*, the latent heat contribution to entropy change shows an initial small increase at $T_C$, followed by a broad plateau and then a sharp decline. In the sample with higher Mn content, E, the plateau is lower in absolute value and shows a similar sharp decline. In the highest Mn content sample (B) only the sharp decline remains.

Figure 5 shows the direct influence of hydrogenation on sample B. Interestingly the magnitude of the latent heat at $T_C$, is comparable in the two sample states. The sharp decay is exponential in both cases, showing that there is a region approaching the $T_{crit}$ where the evolution is a thermally driven process and, 10 times faster in the hydrogenated state at higher *T* (see figure 5).

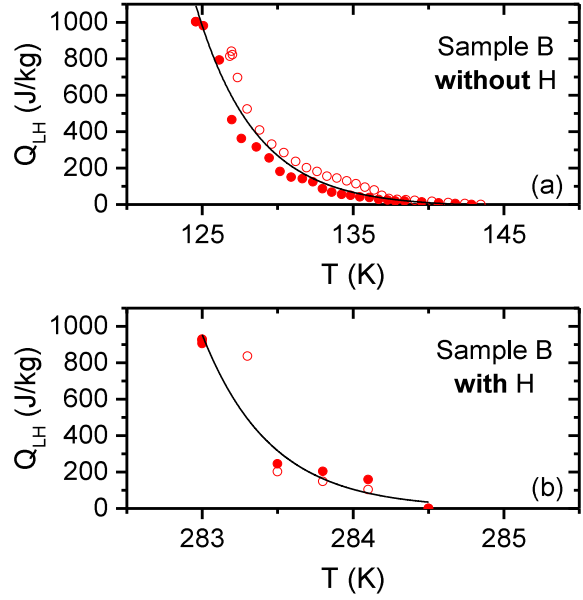

FIG. 5. The latent heat measured in sample B (a) without hydrogen and (b) with hydrogen. Solid symbols correspond to cooling/field application, open symbols correspond to heating/field removal. The lines correspond to an exponential decay fit. The decay is approximately 10 times faster in the hydrogenated state.

Rather than use ($T_C$ - $T_{crit}$) as a measure of the first order character we are interested to introduce a generic model independent parameter $\Omega(B)$ which could in principle be used to compare between materials families:

$$\Omega(B) = \frac{1}{T_2-T_1}\int_{T_1}^{T_2}\left[1-\left(\frac{\Delta S_{Maxwell}-\Delta S_{CC}}{\Delta S_{CC}}\right)\right]_B dT, \qquad (7)$$

where $T_1$ (low) and $T_2$ (high) are temperatures well into the ferromagnetic and paramagnetic states of the material, respectively, and $B$ is the maximum of the field range.

We use the latent heat determined estimation of the Clausius-Clapeyron component $\Delta S_{CC}$ and the magnetization determined value of the total entropy $\Delta S_{Maxwell}$ to obtain values for $\Omega$(1.5 T) = 0.73, 0.37 and 0.16 for samples G, E and B, respectively. Further to this we investigate whether the same information that we have gathered from the latent heat can be extracted directly from magnetization-field curves using the Clausius-Clapeyron (CC) equation (2). Figure 6(a) shows the phase line derivative and the estimated $\Delta M$. As discussed above, it is usually observed that the slope of the phase line is constant or varies only very little. This is not the case in the samples studied here as shown in figure 3(c) and repeated here for samples B, E, G so as to make a direct comparison to $\Delta M$. The slope of the phase line varies significantly in a trend opposing the change in magnetization (resulting in a plateau in $\Delta S_{CC}$) and helps to explain the functional form of the directly measured latent heat. In sample G, the product of the two terms results in an initial increase in the $\Delta S_{CC}$ estimate when dominated by the changes in phase line slope, followed by a decrease where the decreasing change in magnetization dominates. Although the $\Delta S_{CC}$ taken from the magnetization data reproduces the functional form of the directly measured latent heat contribution, as shown in figure 6(b), the magnitude estimated from the CC equation is significantly larger. We find that the closest agreement can be realized by a) using fragment samples for both types of measurement and b) defining the $\Delta M$ change by performing minor M-H hysteresis loops (in increasing and decreasing fields), to establish the precise field and corresponding M at which irreversibility (hysteresis) sets in. These additional measurements are indicated in figure 6(b). Previously a fitting routine was used to extract ΔM to perform an estimate of the first order contribution to the total entropy change, but here too the difficulty in direct extraction once the transition became only weakly first order, was discussed.[36] These issues set out the limitation of the Ω parameter.

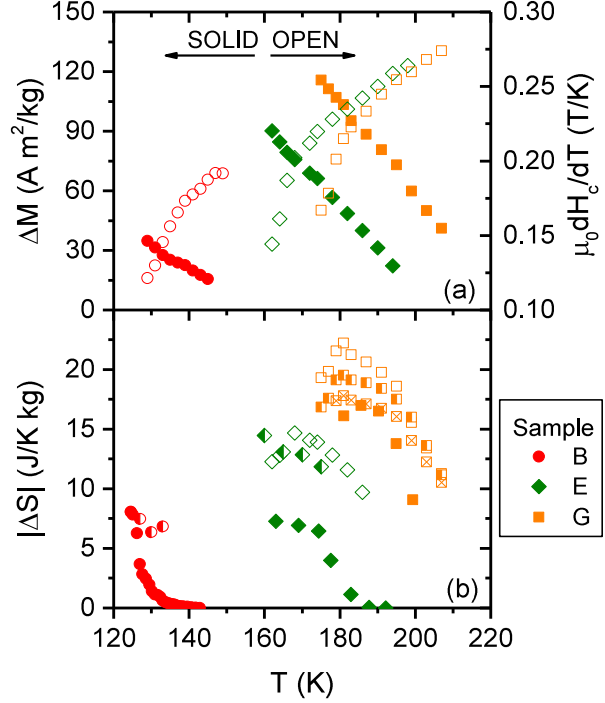

FIG. 6. (a) ΔM (taken from the onset of hysteresis) and $\mu_0 dH_C/dT$ variation as a function of temperature taken from bulk samples. (b) Directly measured latent heat (solid symbols) compared with the Clausius-Clapeyron equation using ΔM taken from the onset of hysteresis (open and half-filled symbols for bulk and fragment data, respectively) and from minor hysteresis loops on a fragment of sample G (crossed squares). All data correspond to magnetic field application/cooling.

An interesting observation in these samples is the large changes in phase line slope which are unusual. They appear related to a secondary non-field-driven phase transition above the $T_C$ which is present in the heat capacity data shown in figure 7. Once the FOPT moves beyond this peak the phase line slope seems to fall on a universal line across the series. The fact that the phase transition at $T^*$ exists at low magnetic fields, but as the field is increased and the FM transition moves to higher temperature, the feature is incorporated into the main ferromagnetic transition, as shown in figure 7(b), suggests the peak is of magnetic origin. The change of slope of the $\mu_0 dH_C/dT$ at the temperature where the $T_C$ and $T^*$ coincide supports this statement. Recently,[8] it was shown from density functional theory that Mn adds antiferromagnetically into the La(Fe, Si)$_{13}$ lattice, and hence it is tempting to attribute this feature to an AFM ordering of regions of the sample where the Mn resides. We also speculate that the $T_N$ of these regions is influenced by the size of the region, as the transition appears to broaden for samples with more Mn overall. Although this is a low field oddity, only observed due to the sensitivity of our calorimeter, the feature appears to affect the shape of the $\mu_0 dH_C/dT$ and therefore has some influence on the overall magnetocaloric entropy change.

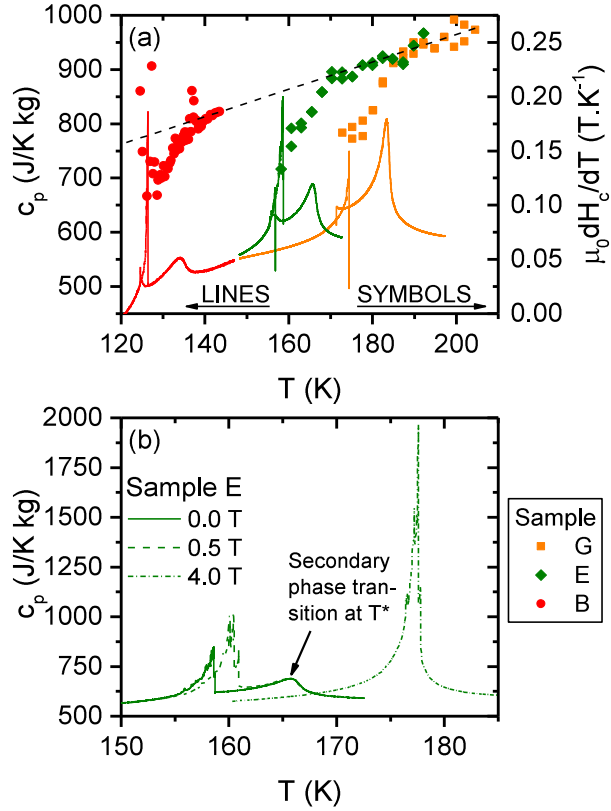

FIG. 7. (a) Ac heat capacity of the dehydrogenated samples in zero magnetic field shows a peak in heat capacity above the FOPT. The $\mu_0 dH_C/dT$ derived from the measurements on the same fragments are also shown. (b) The peak is not affected by small field, however, it does not exist in the FM state confirming that it is associated with the primary phase. The sharp increase in $\mu_0 dH_C/dT$ is associated with the absorption of this peak as the FOPT is shifted to higher temperatures in field.

## IV. SUMMARY

We have studied a series of Mn doped La(Fe, Si)$_{13}$ samples to examine the explicit change of first order character in the presence of magnetic field and temperature. We show the dramatic change of the first and second order character of the transition when the samples are hydrogenated. Remarkably, although the character of the transition is an important property for the development of the field, no one simple model exists to identify the nature of the transition across different material families and the defining sharp features are usually broadened by material inhomogeneity. Using a direct method we have measured the latent heat of the transition, and introduce a new model independent parameter to define the degree of first order character explicitly. The use of the parameter will allow different material families to be compared directly. In addition, we find an interesting feature in the heat capacity associated with the presence of Mn in the samples.

**Acknowledgements**

This work is in part funded by the European Community's 7th Framework Programme under Grant agreement 310748 "DRREAM", and the UK EPSRC rant number EP/G060940/1